%% file: paper.tex
\documentclass{content-experiences/sig-alternate}

\usepackage{graphicx,lipsum,url}
\usepackage{pifont}
\usepackage{color}
\usepackage{caption}
\usepackage{subcaption}
\usepackage{comment}
\usepackage{soul}

\begin{document}

\title{Technical Report: Design, Implementation and Deployment of Intermittency-aware Cellular Edge Services for Rural areas}
\numberofauthors{1} 

\author{
{\rm Talal Ahmad}\\
\and
{\rm Alfred Afutu}\\
\and
{\rm Kessir Adjaho}\\
\and
{\rm Yaw Nyarko}\\
\and
{\rm Lakshminarayanan Subramanian}\\
}
\maketitle

\input{content-experiences/abstract}
\input{content-experiences/1-intro}
\input{content-experiences/relatedwork.tex}

\input{content-experiences/3-arch-new}

\input{content-experiences/4-archspecs}
\input{content-experiences/5-experiences.tex}

\input{content-experiences/conclusion.tex}

\bibliographystyle{abbrv}
\bibliography{references}

\end{document}

%% file: content-experiences/abstract.tex
\begin{abstract}

Rural areas continue to be plagued by limited data connectivity due to poor and unreliable power infrastructure, poor backhaul connectivity and lack of economic incentives for telecom providers. These factors continue to be an impediment to providing reliable, highly available and bandwidth intensive mobile services and applications to function well in such contexts. This paper presents the design, implementation and deployment of {\em GreenLinks}, a new ground-up platform that enables intermittency-aware, reliable and available cellular services and application support in rural contexts under extreme operational environments with limited power and no existing cellular coverage. Unlike the conventional monolithic cellular network architecture, GreenLinks enables a distributed (and potentially disjoint) collection of open, programmable cellular base stations to
offer a gamut of both conventional cellular services and new forms of distributed edge services. GreenLinks offers new application primitives to support different types of intermittency-aware distributed mobile edge services including P2P transactions, voice-based social media applications, and distributed mobile sensing applications. Using a participatory sensing white spaces approach, a GreenLinks base station can co-exist with other conventional cellular
networks. 


\end{abstract}

%% file: content-experiences/1-intro.tex
\section{Introduction}

Cellular networks have achieved significant penetration levels in developing regions in the past decade. However, much of this growth has been predominantly centered in urban regions, with relatively low rural presence~\cite{wildnet,2P,litmac,delaycheck,beyond,kurtis_ictd}. The existing cellular connectivity model is not economically viable for rural settings due to the following fundamental challenges in rural contexts:

\noindent{\bf Power:} Rural regions lack stable and reliable grid power. Cellular networks are inherently power hungry in rural settings (to blanket large areas) and often rely on diesel-powered generators for power, which is a highly expensive proposition~\cite{ahmad2015solar}.

\noindent{\bf Lack of reliable backhauls and data services:} Data connectivity is largely absent in rural settings; where connectivity is available, options are generally restricted to voice and SMS messages~\cite{hermes, smarttrack}. Backhaul has been a roadblock to the introduction of new data-driven services.

{\bf Reliability:} Maintenance of rural wireless networks is very hard due to a myriad of reliability problems, power-related problems and lack of local expertise to fix them. Lack of clean power is known to trigger frequent device failures in rural networks~\cite{beyond}.

In this paper, we present the design, implementation and deployment of {\em GreenLinks}, a new ground-up cellular network that enables basic communication and provides support for intermittency-aware edge based services in rural contexts. The basic building of the GreenLinks
network is a Virtual Cellular Node (VCN) which are open, programmable cellular base station platforms~\cite{range, openbts, openradio, softcell} that enable easy deployment of low-power software defined base stations that function effectively over an IP backplane. In addition, each VCN requires access to a backhaul link and needs to operate in licensed spectrum band without owning any spectrum. GreenLinks enables a distributed collection of VCNs to inter-operate
as a single cellular network that can offer reliable, available and intermittency-aware cellular edge services to mobile devices. GreenLinks provides a set of identity management and distributed mobile service primitives that enables conventional cellular services and new forms of distributed edge services and mobile applications. This paper makes the following contributions:

{\em 1. Whitespaces detection:} GreenLinks reuses spectrum allocated to cellular operators using a participatory sensing whitespaces approach to determine unused channels in order to work without interference with conventional cellular operators. 

{\em 2. Reliable Backhauls:} In areas which are at the boundaries of conventional cellular coverage (where signals are negligible, but existent), we demonstrate how one can use a simple directional signal booster to build a reliable backhaul abstraction using parallel boosted data links between a VCN and a conventional cellular tower.

{\em 3. Identity Management:} To provide Internet and conventional cellular services, GreenLinks supports an identity management layer that different users across VCNs to join the network without requiring any modifications to the end-hosts. 

{\em 4. Distributed Mobile Edge Services:} Given the edge programmability, GreenLinks provides application primitives to enable new forms of intermittency-aware, cloud-controlled, distributed, mobile edge services where every VCN maintains critical application state to enable mobile services to function in an intermittency-aware manner in the face of unreliable backhauls and network failures. These services are also {\em distributed} in the sense that users across VCNs can cooperate and run different types of peer-to-peer applications and community-specific applications. 


We demonstrate the effectiveness of the GreenLinks model using a real-world 3-VCN deployment including a solar-powered cell tower installation in rural Ghana. Our Ghana installation also supports different types of distributed mobile services (as outlined earlier) that have been used by rural users within the community. In summary, we believe that GreenLinks represents a change in paradigm in how we design cellular networks, and could enable ground-up innovation where several budding rural entrepreneurs could launch competing cellular network services in rural localities while seamlessly co-existing with existing operators.
 

%% file: content-experiences/relatedwork.tex
\section{Related Work}

{\bf Software-based microcells:} 
Recent advances in hardware and open-source software has made available inexpensive cellular equipment broadly accessible. For example, OpenBTS is an open-source GSM base transceiver station (BTS) implementation which has enabled a wide range of projects aimed towards building small-scale VCNs. Heimerl et al.\cite{kurtis_ictd} demonstrated the viability of independently run, locally operated cellular networks. Similarly, Rhizomatica\cite{rhizo} has deployed several community-run cellular networks in Oaxaca, Mexico with a short-term experimental spectrum license. Zheleva et al.~\cite{zheleva} deployed a similar system for purely local communications in Zambia. 

\textbf{Backhauls and rural mesh networks:} The design of GreenLinks builds upon a large body of work across rural wireless networks including Fractel~\cite{fractel}, Digital Gangetic Plains~\cite{inside_out,gangetic,2P}, WiLDNet~\cite{wildnet,beyond}. There has been immense around design of various different Line of Sight(LOS) and Non Line of Sight Cellular(NLOS) backhauls~\cite{tarana,kumu}.   

\textbf{Cognitive Radio and GSM whitespaces:} Most of the literature on whitespaces is in the space focuses on TV spectrum, our work is more closely related to work on re-use of cellular spectrum. Sankaranarayanan et al.~\cite{sankaranarayanan2005bandwidth} propose reusing timeslots in a GSM cell for adhoc networks during periods when the GSM cell is lightly utilized. Buddhikot et al.~\cite{buddhikot2009ultra} describe a system for indoor femtocells to dynamically share spectrum with incumbent carriers by operating over ultra wide bands. Yin et al.~\cite{yin_beijing} proposes a similar system and provides measurement results which indicate that unused spectrum (i.e., whitespace) exists even in a dense, urban environment (Beijing). In contrast to these, we focus on reusing GSM whitespaces to provide GSM service by means of macrocells. The only work that recycles GSM whitespaces for reuse by a VCN is NGSM~\cite{gsmws}. Later in this paper we highlight how our approach for doing GSM whitespaces builds on top of NGSM and performs better.

\textbf{Intermittency-aware networks:} Dealing with intermittent failures has been studied in the context of Delay Tolerant Networks~\cite{dtn_routing,seth2006lcc}.  One of the first real world DTN deployments was DakNet which provides low-cost digital communication to rural areas using physical transportation links~\cite{pentland2004daknet}.  There have been several routing DTN-based routing algorithms like Encounter based
routing~\cite{nelson2009encounter} and MobySpace~\cite{leguay2005evaluating}. \cite{burgess2006maxprop, juang2002energy, chaintreau2007impact} are other works that use DTN centric ``store and forward'' algorithms to progressively forward
messages until they reach their destination.
The intermittency-awareness abstraction of our solution is fundamentally different from the traditional DTN abstractions since GreenLinks is primarily tailored and designed for the cellular network context and primarily focuses on higher application-centric ``store and forward'' abstractions; the DTN routes for messages are preset by the network structure which changes infrequently.

%% file: content-experiences/3-arch-new.tex
\section{GreenLinks}

\begin{figure}
\centering
{
	\includegraphics[scale=0.4]{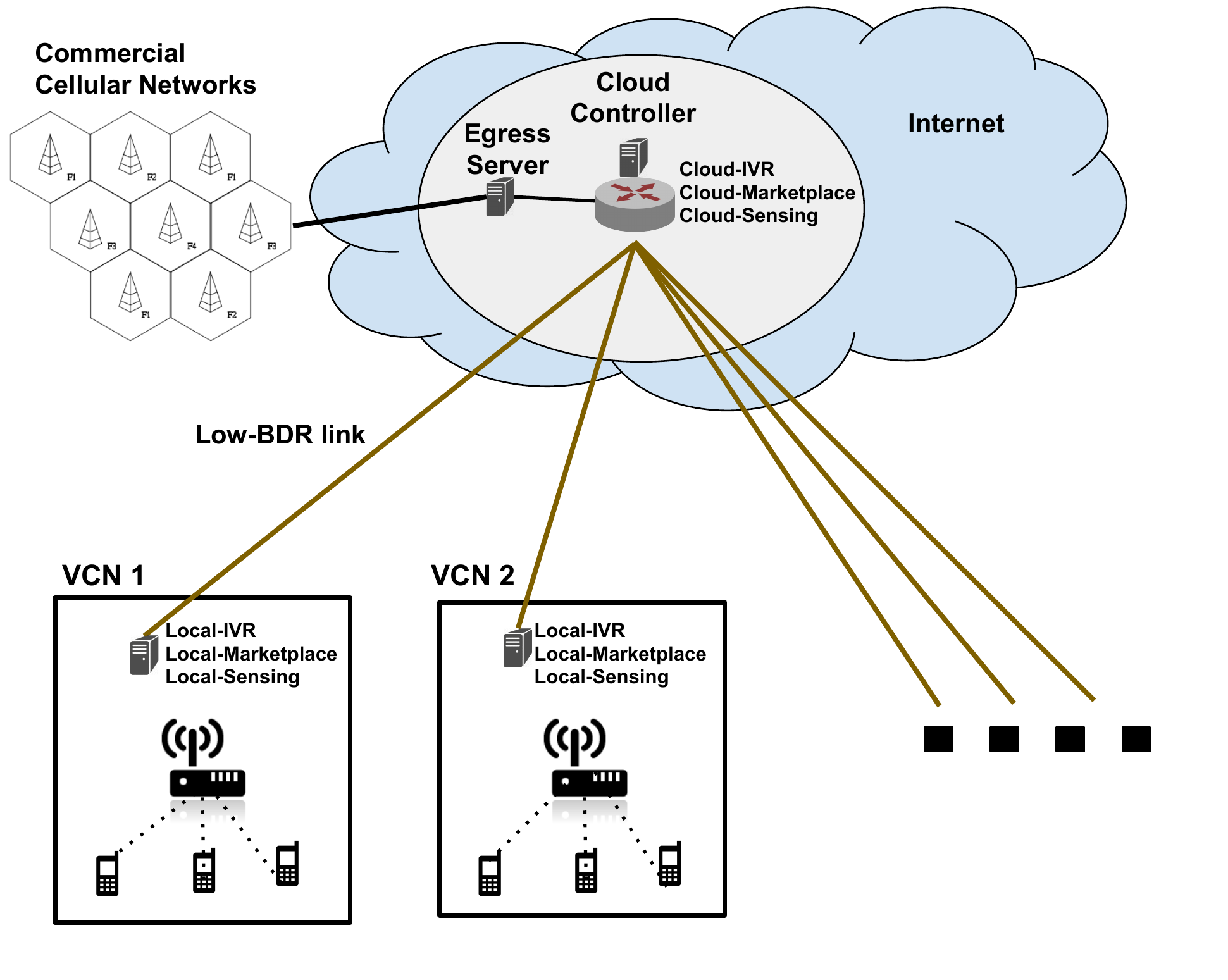}
}
\caption{GreenLinks System Architecture}
\label{fig:systemArch}
\end{figure}

In this section, we describe the GreenLinks architecture and how it
enables new forms of intermittency-aware, distributed cellular edge
services.  The GreenLinks architecture is shown in
Figure~\ref{fig:systemArch}.  The basic building block of GreenLinks
are Virtual Cellular Nodes (VCNs) which leverage open, programmable
cellular basestations with local compute capabilities. GreenLinks aims
to connect multiple VCNs which can be located in different geographic
localities to cohesively behave as a single cellular
provider. GreenLinks is specifically designed for rural contexts with
limited or no existing cellular coverage and we assume that each VCN
operates with a backhaul link with highly constrained bandwidth and high latency. We define such backhaul links as low
Bandwidth-Delay Ratio (BDR) links.  Examples of such backhaul links
are: existing 3G coverage with weak signals, low bandwidth satellite
links, microwave links and long distance Wi-Fi links. The backhauls
can fail arbitrarily due to loss in power or other exogenous
factors. 

GreenLinks offers a combination of conventional cellular services
(voice calls, messaging, voice messages and data services) and new
forms of intermittency aware distributed applications.  Despite constant
power failures and network failures which may be common in rural
contexts, GreenLinks aims to provide high availability of basic
services and high interactivity for distributed applications. To
achieve this, GreenLinks offers an intermittency aware service
abstraction where every cloud service or application offered to mobile
users has a cloud instance and several edge instances (also referred
as local instance).  The local instance of each application runs at
the compute server attached to the base station, while the cloud
instance runs in a data center.  We partition each application in such
a way that most of the application processing is done at the local
component. In addition, each application needs an intermittency-aware
synchronization mechanism between the local instances running across
VCNs and the cloud instance to ensure the consistency or availability
guarantees required for various functions within the application.
Figure~\ref{fig:systemArch} outlines three different example
distributed applications that we have built on top of the GreenLinks
service abstraction: (a) a Marketplace application for supporting P2P
transactions across users; (b) an IVR based social media application;
(c) distributed data gathering and sensing. We describe these
applications in greater detail in the paper.  We also have an {\em Egress server} which is responsible for communication between VCNs and the conventional cellular networks.

\subsection{Our VCN deployment}
\label{ghana}

\begin{figure}
\centering
{
	\includegraphics[scale=0.07]{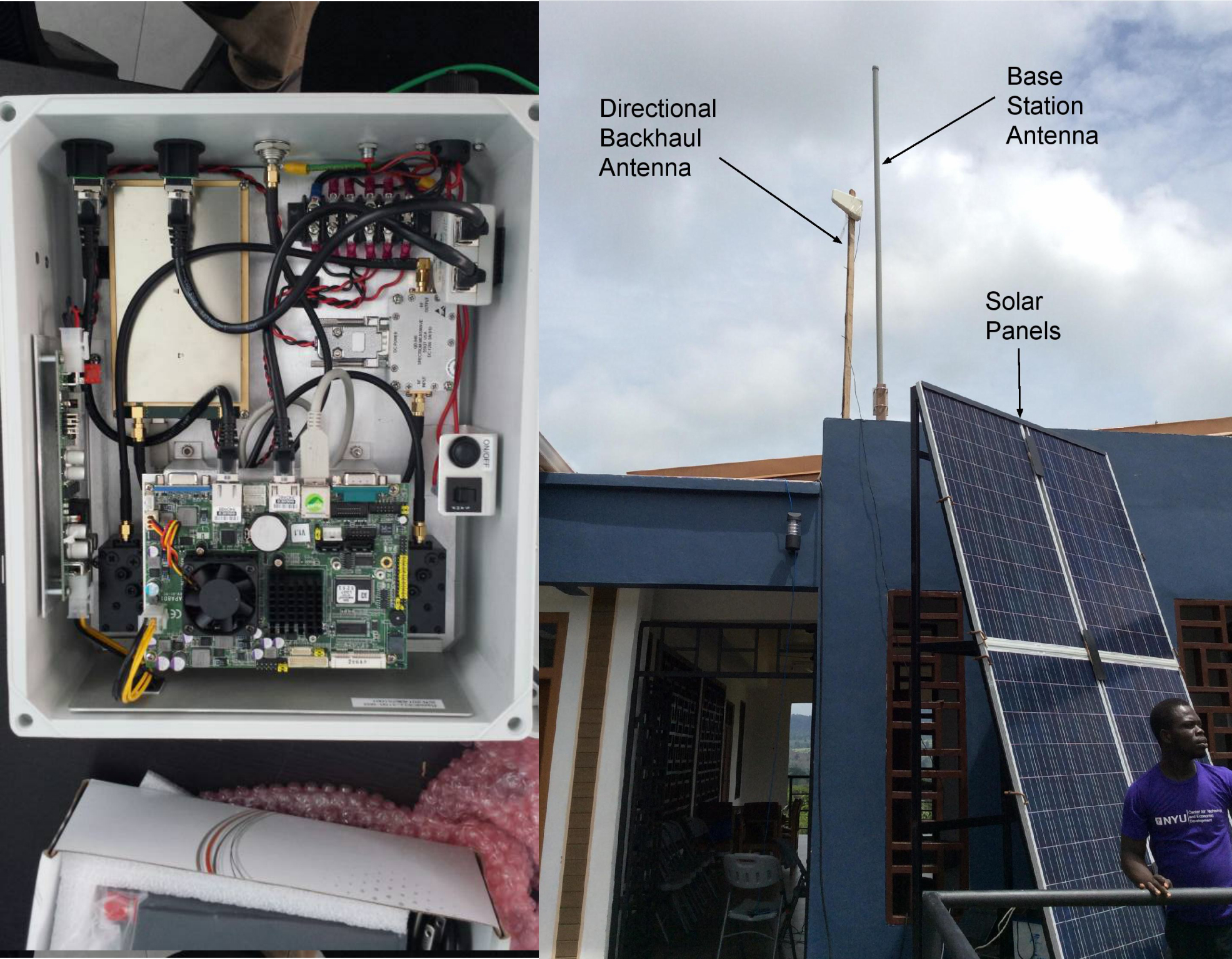}
}
\caption{Our VCN setup in Kumawu, Ghana}
\label{fig:ghanaVCN}
\end{figure}

Our deployment consisted of three VCNs located in the USA, Ghana and
UAE respectively, with one centralized controller located in
Germany. Figure~\ref{fig:ghanaVCN} shows the deployment we have in
Ghana, at the edge of existing cellular coverage by commercial
telcos. Pictured on the left side of Figure~\ref{fig:ghanaVCN} is the
cellular base station hardware used, a RapidCell manufactured by Range
Networks\cite{range}. To the right side of Figure~\ref{fig:ghanaVCN}
is the backhaul and base station antennae, and the solar panels used
to power the complete setup. We used this VCN to extend coverage
beyond the existing coverage area, with the backhaul comprising of
analog signal booster with a directional antenna (facing the 3G base
station) to boost the weak 3G signal to create a backhaul for the
VCN. We also deployed a whitespaces solution to make sure that we do
not interfere with the primary operators in the area while we try to
use the bands in spectrum no one else was using. We also deployed 3
applications on top of the Ghana VCN and used synchronization
primitives to enable applications to run across Ghana and US VCNs. 

We used two hardware platforms as Base Transceiver Station (BTS) nodes. While in Ghana we used a 1 Watt RapidCell from Range Networks~\cite{range}, in USA and UAE we used the 10 mWatt USRP B100 from Ettus Research. Freeswitch PBX was used for routing calls and messages in all VCNs. As the local component of the IVR application, we deployed an Asterisk PBX on compute node to handle IVR calls. The local server was also running an HTTP server for the SMS based and data sensing applications. 


\subsection{Identity Management}

A key building block for supporting different forms of services in
GreenLinks is an intermittency aware identity management
layer. GreenLinks supports three basic types of identities: network
identities, user identities and application identities. In GreenLinks,
the common user identity is the the International Mobile Subscriber
Identity (IMSI) which is physically written on a SIM card and the
network identity is the identity issued by Network against the IMSI,
this is commonly referred to as the phone number of the user.
GreenLinks supports a generalization of the definition of these
network and user identities, in that user identities can encompass
other broad notions of identities and network identities additionally
encompass IP addresses.  Given the basic network setup of a disjoint
collection of VCNs inter-connected over an IP backplane, the cloud
instance of GreenLinks is publicly reachable (with a public IP
address) while each VCN may be positioned behind a NAT. GreenLinks
assumes a flat 2-level hierarchy where each VCN independently
maintains a connection with the cloud and all synchronization and
information routing is performed through the cloud instance. 

The basic identity lookup process translates a user identity to
determine the network identity within the GreenLinks network.
GreenLinks has a lookup server instance within each VCN and cloud
instances. The traditional lookup would typically involve a local
lookup within the VCN and if locally unavailable, a cloud lookup for
the same user identity. The link between the VCN and the cloud has a
low BDR and is typically intermittent. This raises four challenges:
availability, consistency,external names and mobility.
First, to ensure high availability of identity lookup, GreenLinks
supports quick local authentication of user identities, for
pre-registered users and two users within the same VCN can communicate
with each other without the need for backhaul connectivity to the
cloud. A user is pre-registered if the user identity has been
authenticated by the cloud instance and the identity is active in the
local cache. Second, to ensure consistency of identities, local
servers at the VCN cannot assign global user identities but only the
centralized name server (co-located with the cloud controller) issues
identities to all VCNs to avoid naming conflicts. 
Third, the GreenLinks network can operate on two types of names:
unique user identities assigned by GreenLinks or unique user
identities assigned by external trusted entities (cellular identities,
etc.). In the former case, the central name server has complete
control over the user identity space and controls the uniqueness of
GreenLinks assigned names. To enhance availability of names, the
central server can partition the name space and also provide local
VCNs autonomy to assign locally unique user identities.  GreenLinks
assigned names have no semantic meaning outside the GreenLinks
network. In the case of user identities assigned by external trusted
identities, the GreenLinks cloud name server needs to communicate with
the appropriate external entities to verify their authenticity. 
Finally, given the intermittency aware lookup process, GreenLinks is
not best suited to support rapid mobility of user identities across
VCNs. In specific cases, where neighboring VCNs can directly communicate
with each other using a local backhaul, then GreenLinks can support limited
forms of local mobility across VCNs.

GreenLinks also supports distributed applications to define
application specific identities, which refer to unique information
state of the application that may be tied with one or a group of user
identities.  For example, consider the case where two users wish to
perform a BUY/SELL transaction over GreenLinks across multiple VCNs;
each item transacted is associated with a unique application identity
and each buy/sell bid is a mapping between application identities and
user identities. Applications can specify their own consistency,
availability and intermittency-aware tradeoffs across their
application defined identities.  This state is used by the local
instances of the application running at each VCN to selectively
synchronize state with the cloud instance of an application.

GreenLinks also supports intermittency-aware forwarding of state for
applications and user-directed messages. As a simple example, the
messaging application in GreenLinks enables users to leave voice
messages or text messages with a user identity (this can be a
GreenLinks issued identity or even a globally unique mobile number),
which can be locally stored in a VCN and lazily synchronized with the
cloud instance.

\subsection{Distributed Application Primitives}
\label{subsec:prim}

GreenLinks supports a collection of basic application primitives that
enables developers to easily translate existing distributed
applications with minimal changes to operate in intermittency-aware
conditions. The key focus of primitives is to enable an
intermittency-aware key-value store abstraction over the application
identity space that enables applications to specify the right tradeoff
between consistency and availability. Since most VCNs are located in
low-BDR regions, we achieve partial availability by dividing each
application into a larger local component and a smaller cloud
component. The local component of the application is to ensure
significant local processing. This ensures responsiveness (or
application liveness) by keeping local communication and interaction
functional even in the absence of the low-BDR backhaul link. The cloud
component is mainly responsible for saving application data and
facilitating interaction between the local components of various
VCNs. The lack of programmability in conventional cellular networks
forces application deployment extensively in the cloud. This means
that failure of a backhaul link between the base station and core
network inhibits access to even basic applications like calls and
SMS. We discuss some of the primitives we designed and implemented:

\vspace{1mm}
\noindent{\texttt{1. SLOWPUT(Identity i, Type t, Data d)}}: 

\noindent{User \texttt{i} communicates marshalled data \texttt{d} for application \texttt{t}. This primitive is used by the local component for {\em lazy} communication with the cloud component. This operation is lazy because data may not be immediately sent after the primitive is executed; the local server determines the right order to send the data based on bandwidth availability and priority (based on \texttt{t}) of the other requests in the queue. The \texttt{SLOWPUT} primitive is used by applications for which the local component can provide availability at the expense of strong consistency. For example if someone is just try to send an e-mail from VCN $A$ to a user connected to VCN $B$, any inconsistency in the ordering of e-mails at receiver is tolerable.}

\noindent{\texttt{2. FASTGET(Identity i, Type t, Data d)}}: 

\noindent{This primitive is used by the local component of any application to send and receive data from the cloud component in an {\em active} manner. The particulars include a user \texttt{i} communicating data \texttt{d} for application \texttt{t}. This primitive is usually called as a result of an application function when consistency and response time can not be compromised. The communication is not queued, as in \texttt{SLOWPUT} primitive. Instead, this primitive is executed instantaneously, provided the the backhaul link is available. For example, any inconsistencies in the case of a banking transaction are grave, and an strong consistency has to be ensured even if the user does not get a response immediately. Such applications use the \texttt{FASTGET} primitive and marshall the transaction in the data parameter, forwarding it to the central component of the banking application.}

\noindent{\texttt{3. FASTSEARCH(Identity i, Type t, Data d)}}: 

\noindent{This primitive is used by the user \texttt{i} to search for data \texttt{d} of a particular application type \texttt{t}, and returns one or more data points based on search from the cloud component's data store. The communication encapsulated by the \texttt{FASTSEARCH} primitive occurs instantaneously. The \texttt{FASTSEARCH} primitive executes a read operation on the persistent storage of the cloud component, therefore it can be performed even if a \texttt{FASTGET} operation has a write lock. An example use-case of the \texttt{FASTSEARCH} primitive can be to search in the data store of cloud component while a user is on hold with an IVR-based application, and is expecting a response quickly. } 

\subsection{Example Applications}
\label{subsec:example}

In this subsection, we give the examples of distributed applications
that we built using the primitives discussed above. In particular, we
highlight how a functional division is made to identify the functions
that require consistency to those that require availability.

\subsubsection{P2P Transactions}

The application provides a text-based interface (running on top of the SMS service) and enables traders to provide customers the price and quantity of a commodity (usually crops) they wish to sell. It also enables customers to search for a particular commodity, the price at which various traders are selling, and to place a \texttt{BUY} request. The \texttt{BUY} request leads to exchange of contact information between transacting parties, and is logged as a \texttt{SELL}. The text-based interface requires small phrases to commence transactions.

The lazy \texttt{SLOWPUT} is used send the data to the server on sale of a commodity. This is because the order in which \texttt{SELL} request is updated at the cloud does not require strong consistency, but requires availability. Therefore when a user tries to sell, we acknowledge the sell has been received locally(by sending a message back to user) and transfer the sell to cloud in a lazy manner. In case the user is interested in purchasing a commodity, \texttt{FASTSEARCH} is used, and the data returned is sent to the user over one or more text messages. This is because we want strong consistency and do not want to give user a response immediately without ensuring that we are actually able to update the record in the cloud.

\subsubsection{IVR-Based Social Media}

This application enables users to interact with the system by pressing numbers on the phone's keypad. It enables users to leave a message, or to listen to a message left by others. For the Ghana VCN, the application provided an option to either use English or Twi, the local language in the region. The key function is the \texttt{ReceiveHandler} at the local component, which is invoked using a HTTP request on entering the IVR extension. After the initial HTTP request, the call is bridged from base station to the Asterisk IVR server running as the local component of application. The Asterisk server logs each input of user as entered. If the user is recording a message, the \texttt{ReceiveHandler} is executed after its completion. The \texttt{ReceiveHandler} uses the lazy \texttt{SLOWPUT} primitive to send the recorded file to the cloud component because a strong consistency in the ordering of messages received is not required. In case a person wants to listen to a message, the function first checks if the validity of the session. If the session is valid the message is played locally. Otherwise \texttt{FASTSEARCH}  and \texttt{FASTGET} primitives are used to search and fetch the most recent message from the cloud component, while the user is on hold. We use these operations because user is on hold and we want to respond to the user quickly.

\subsubsection{Distributed Data Gathering \& Sensing}

The purpose of the application was to collect data on farms and crops grown. The application provides an interface to create the map of any farm by walking along its edge. This data is sent to the local application component, an HTTP server next to the base station. The local component then uses \texttt{SLOWPUT} to upload this data to the cloud component. Our choice in primitive is valid as the application does not require strong consistency and we can afford to have unordered data at the cloud component. This application is relatively simple because data only needs to be uploaded to the cloud component of the application, and there is no need for downloading data.

%% file: content-experiences/4-archspecs.tex
\section{Operational Challenges}

\subsection{Detecting Whitespaces}	
\label{sec:gsm}

GSM spectrum license is required for operating in a particular area. Most countries issue licenses to a handful of national carriers and have no regulatory framework for smaller providers. For our VCN, we had no spectrum allocated to operate. This is true for most community based cellular deployments \cite{gsmws,rhizo}. Though licensed in its entirety to primary operators, majority of the cellular channels remain unused in rural localities. These unused bands of spectrum are called GSM whitespaces. In order to identify and reuse these whitespaces without any collaboration from primary providers, we build a sensing-based participatory model for collecting channel energy information in the locality. 

\begin{figure}
\centering
       \includegraphics[scale=0.22]{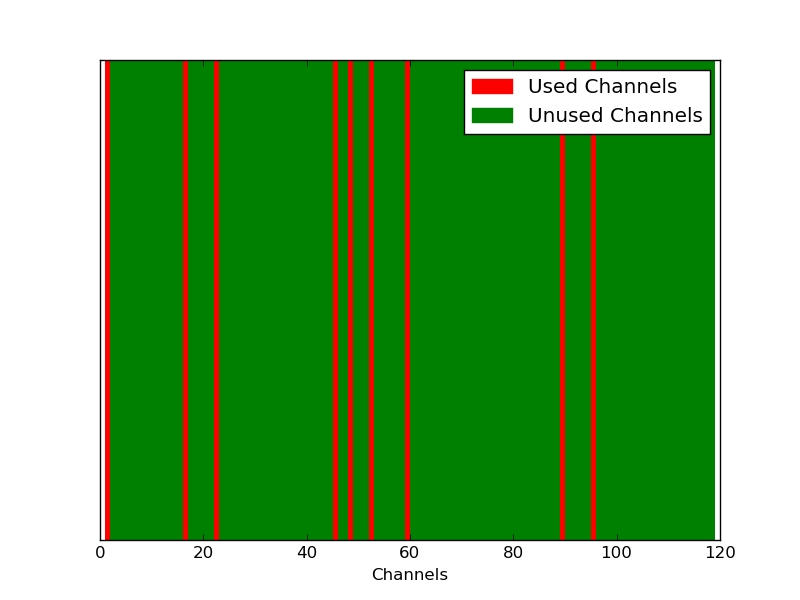}
\caption{Used and unused channels over one week in Kumawu Ghana.}
\label{spectrum}
\end{figure}

Our solution takes advantage of the fact that GSM handsets continually measure spectrum occupancy in their proximity and report these measurements to their serving cell. Phones do this to enable the serving cell to negotiate a hand-over with a cell tower that has a stronger signal. We utilize this functionality and provide cellular devices with channels to scan. To phones, these are the channels on which neighboring base stations (of our cellular network) are operating - so they look for energy on those channels and report it. Although this approach is similar to NGSM~\cite{gsmws}, we make the following improvements: (i) We do not necessitate two different base stations. Instead, we wait for the connected calls to complete in case it is time to shift to a different channel, (ii) We increase transmission power periodically based on sensory feedback from phones to slowly expand coverage, unlike NGSM which starts at full power on a random channel and has a greater possibility of interfering with primary, and (iii) We have a voluntary model where we incentivize volunteers to install an application that periodically makes calls and sends messages to the base station thus ensuring periodic stream of measurement reports even in low activity periods (discussed later in the section).

\begin{figure}
\centering
\begin{subfigure}[b]{0.23\textwidth}
       \includegraphics[scale=0.22]{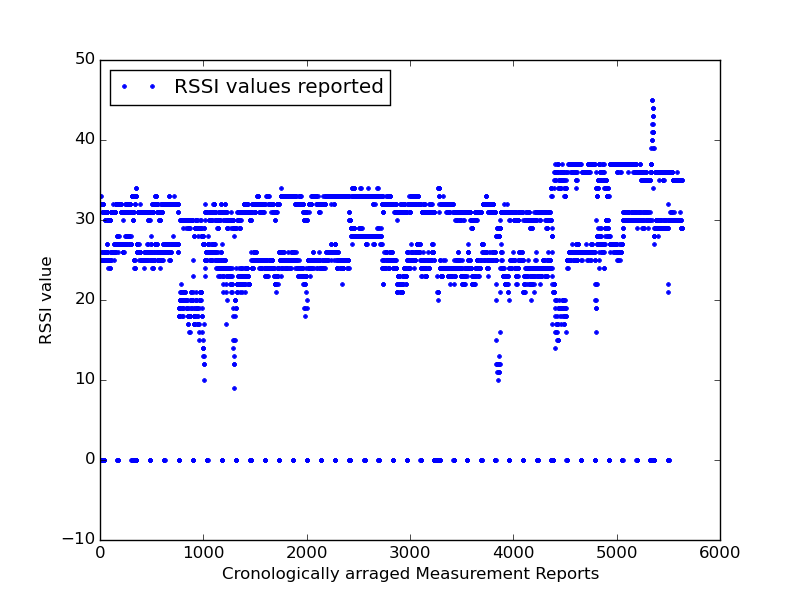}
	\caption{Channel 45 energy}
	\label{channel_value45}
\end{subfigure}
\begin{subfigure}[b]{0.23\textwidth}
       \includegraphics[scale=0.23]{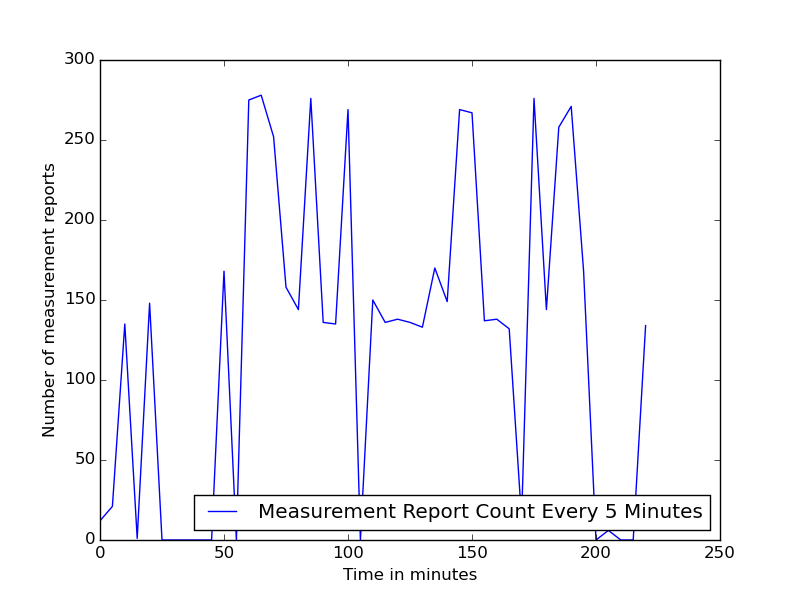}
	\caption{Channel 45 frequency}
	\label{channel_frequency45}
\end{subfigure}
\begin{subfigure}[b]{0.23\textwidth}
       \includegraphics[scale=0.22]{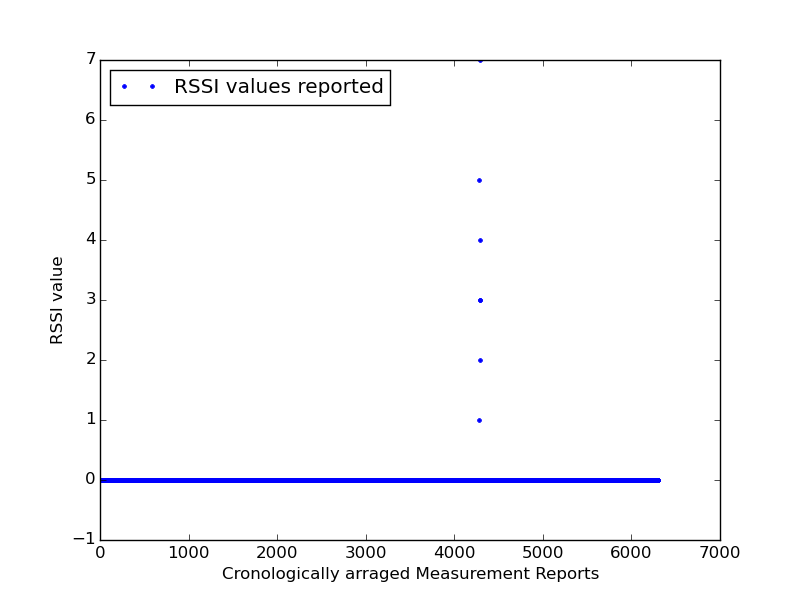}
	\caption{Channel 48 energy}
	\label{channel_value48}
\end{subfigure}
\begin{subfigure}[b]{0.23\textwidth}
       \includegraphics[scale=0.23]{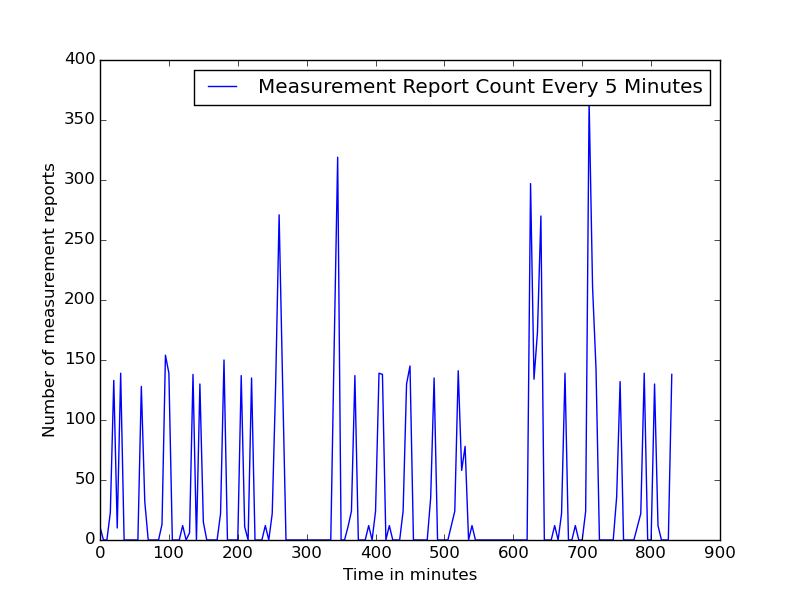}
	\caption{Channel 48 frequency}
	\label{channel_frequency48}
\end{subfigure}
\caption{Channel energy as reported by various measurement reports and frequency of reports at a 5 minute granularity.}
\label{channel_info}
\end{figure}

We modified the peering component of OpenBTS to add fake neighbors, and add the corresponding channel that we intend to scan to each fake neighbor. We stop OpenBTS from validating the fake neighbors using an IP-based protocol. Once the fake neighbors are successfully set, we wait for the base station to send the new {\em neighbor list} out to the phones. Once the phones get the neighbor list, they start reporting the energy on the fake neighbor channels. We constantly run \texttt{tcpdump} to collect \texttt{MeasurementReport} packets being exchanged between phones and the base stations. For this, we modify the code base~\cite{gsmws_code} made available by the developers of NGSM.  In Figure~\ref{spectrum}, we present the detected whitespaces over the period of one week in 124 channels of the P-GSM-900 band. It can be observed that only 9 channels out of 120 were being used, while the rest of the channels were unoccupied. 

Apart from this we also developed an Android application that was given to a few volunteers. This application helped us gather reports frequently by sending SMS between volunteers and the base station. More communication by phones guarantees that regular reports are exchanged. Figure~\ref{channel_value45} shows, in chronological order, the various readings over a period of approximately 4 hours. Figure~\ref{channel_frequency45} shows, for the same channel, the plot of how many reports were received every 5 minutes. These two graphs highlight the fact that there is a constant need for sensing the channel because even for channels clearly being used by a neighboring base station, there are a significant number of reports suggesting that the energy is zero (the horizontal line at y=0 in Figure~\ref{channel_value45}). Taking a closer look at reports pertaining to channel 48 in Figure~\ref{channel_value48} and Figure~\ref{channel_frequency48}, we observed that it can take thousands of zero energy packets and several minutes before a packet reporting a positive energy on a channel is received. One possible reason explaining this phenomenon is that the overlap radius is very small and only a few mobile devices (maybe single device) were able to see energy on a channel and for a very small amount of time. Because of this, we argue for having some phones as volunteers in the area who have our application which keeps sending SMS in the background to other volunteers and the base station and produces regular report. In the case of channel 45, we have a relatively constant stream of packets across time as seen in Figure~\ref{channel_frequency45} but while in case of channel 48 in Figure~\ref{channel_frequency48} there are several minute gaps where no report is received about the channel. This highlights the need for having more volunteers; we add multiple fake neighbors and there is no way of controlling a single channel from which we receive reports.

\subsection{Boosted Cellular Backhaul}
\label{sec:backhaul}
\begin{figure}
\centering
{
	\includegraphics[scale=0.15]{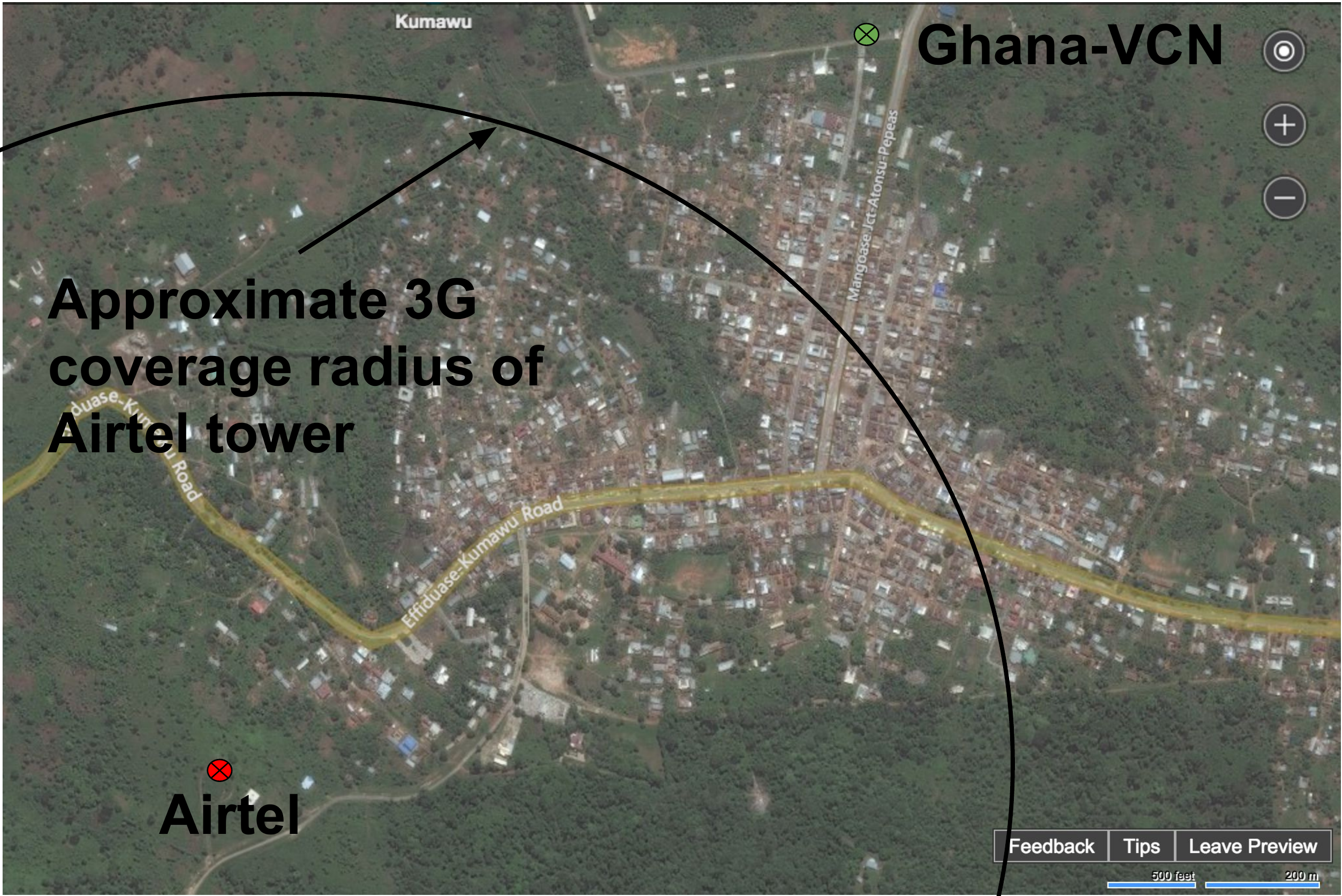}
}
\caption{Location of the Ghana VCN and nearby conventional tower}
\label{fig:ghanaVCN_location}
\end{figure}

\begin{figure}
\centering
\begin{subfigure}[b]{0.25\textwidth}
       \includegraphics[scale=0.22]{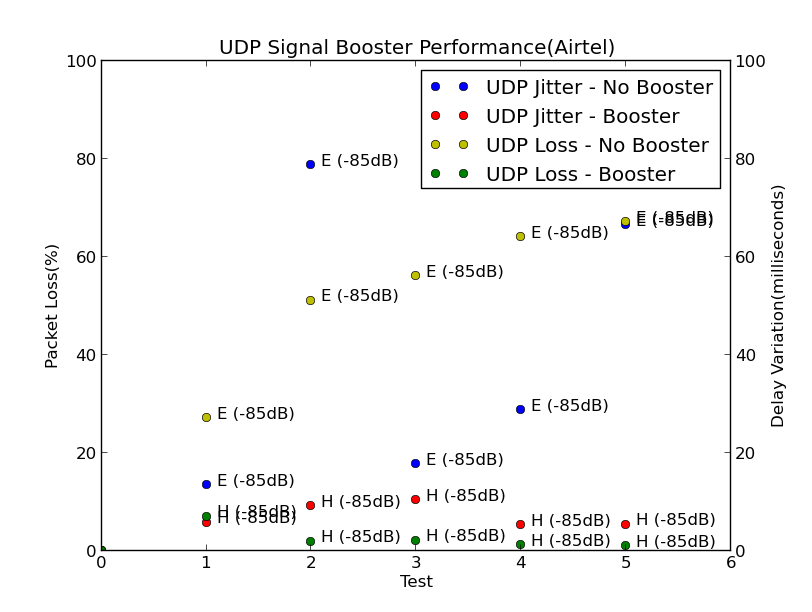}
	\caption{UDP}
	\label{bandwidth_udp}
\end{subfigure}
\begin{subfigure}[b]{0.22\textwidth}
       \includegraphics[scale=0.22]{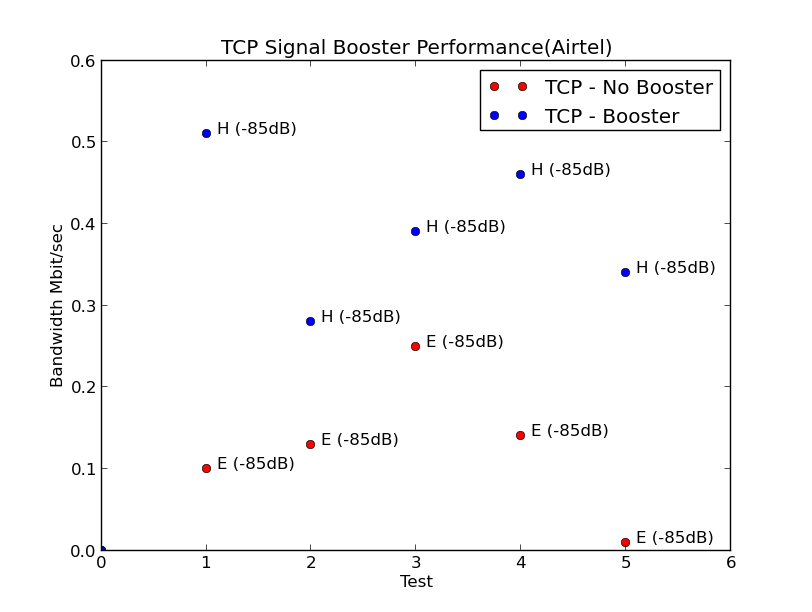}
	\caption{TCP}
	\label{bandwidth_tcp}
\end{subfigure}
\caption{Performance of Airtel connection with and without boosted backhaul}
\label{airtel}
\end{figure}

Designing reliable backhauls in rural contexts is a challenging proposition. We outline a simple and potentially powerful approach for building reliable backhauls using directional signal boosters coupled with parallel cellular connections. Specifically in areas which are at the edge of cellular connectivity, we believe that the parallel boosted links represents an easy-to-deploy solution for building reliable backhauls. 

In Figure~\ref{fig:ghanaVCN}, we have pointed out the backhaul's antenna, which is a high gain directional receiving antenna pointing towards the 3G base station. The rest of the backhaul setup consists of a signal repeater attached to an indoor omni-directional antenna. The weak signal is received at the directional antenna and is fed to the repeater using coaxial cables. The output from the repeater is transferred to the indoor antenna, again using coaxial cables. The key challenge stems from the fact that the Ghana-VCN was located at the edge of coverage as shown in Figure~\ref{fig:ghanaVCN_location}. By mounting the receiving antenna high enough, we were able to get a decent boosted 3G signal. We had multiple phones containing SIM cards belonging to the Airtel Network next to the indoor antenna. We tethered these phones to our setup and had multiple parallel boosted links as backhauls. It is important to note that we were able to obtain 3 boosted links with a single signal booster. 

We then used approaches like Ubuntu Bonding~\cite{bonding} to use the multiple backhauls together in various configurations like active backup (where one link is used and the other is backup) and load balancing (where outgoing traffic is divided into the available links equally). However, the major limitation of Ubuntu Bonding is its inability to divide a single flow into multiple subflows. Multi-path TCP~\cite{raiciu2012hard} is another option that can help aggregate multiple links, but it requires changes to the Linux kernel and some support from middleboxes. 


We used Phonetone K3B0918217CLB kit which comprised of a 10 db directional receiving antenna, 9db indoor panel antenna, and 70db analogue booster capable of working in 900/1800/2100 MHz GSM bands. We were able to capture the 3G signal from Airtel which was not available earlier in the vicinity of VCN. In Figure~\ref{airtel}, we present the performance of the signal booster based backhaul in Ghana, where the bandwidth gain in TCP, and packet loss and jitter improvement in the case of UDP (both measured using \texttt{iperf}) is plotted for a single boosted link against trials. The parentheses next to each point in the graph contains the average measured \texttt{RSSI} value. This value was measured using Android phones that utilized the boosted backhaul to the BTS node. The effect of signal boosting of the Airtel network is evident in Figure~\ref{airtel}; we were able to change the transmission rates from EDGE (denoted by E in graph) to that of HSDPA (denoted by H in graph). Consequently, we were able to get significantly more bandwidth as reported by Figure \ref{bandwidth_tcp}. We also reduced the packet loss rate and jitter in UDP, as reported by Figure~\ref{bandwidth_udp}.

%% file: content-experiences/5-experiences.tex
\section{Application Experiences}
\label{sec:implementation}

All the three distributed applications we developed have been deployed
and tested in the local rural community in Kumawu, Ghana. We outline
some of the technical experiences in running these applications on the
wild. Due to space considerations, we primarily focus on the
marketplace P2P transaction application.  

{\bf Marketplace application:} To test the local and cloud
component of the Marketplace application, we built an Android
application that enables phones to automatically \texttt{SELL} and
\texttt{BUY} items in a periodic manner. We had 4 phones sending
\texttt{SELL} messages and 3 phones sending \texttt{BUY} and
\texttt{SEARCH} messages. 5 of the 7 phones were connected to the BTS
in Ghana and the others were connected to the BTS in New
York. Simultaneous with the above application we made calls manually
and a script running at local server generated data to simulate the
marketplace server receiving data. The node in UAE did not participate
in these measurements.


\begin{figure}
\centering
\begin{subfigure}[b]{0.23\textwidth}
   \includegraphics[scale=0.16]{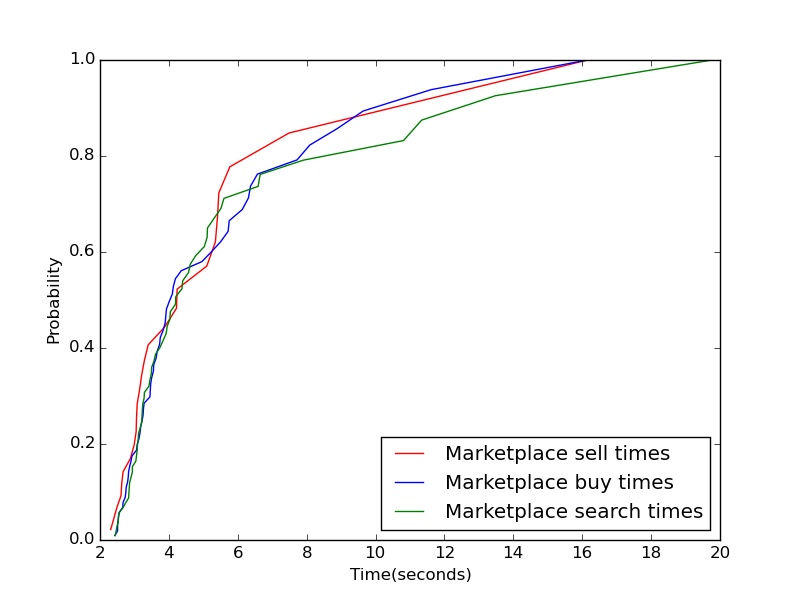}
	\caption{Ghana}
	\label{ghana_cdf}
\end{subfigure}
\begin{subfigure}[b]{0.23\textwidth}
	\includegraphics[scale=0.16]{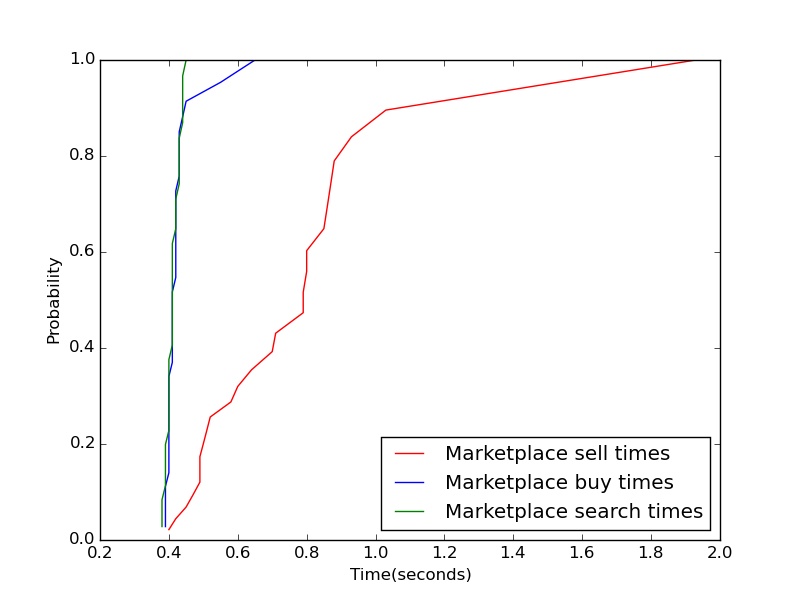}
	\caption{New York}
	\label{ny_cdf}	
\end{subfigure}
\caption{CDF of time it takes for the different functions in the P2P Marketplace application}
\label{cdf}
\end{figure}
\begin{figure}
\centering
\begin{subfigure}[b]{0.23\textwidth}
   \includegraphics[scale=0.16]{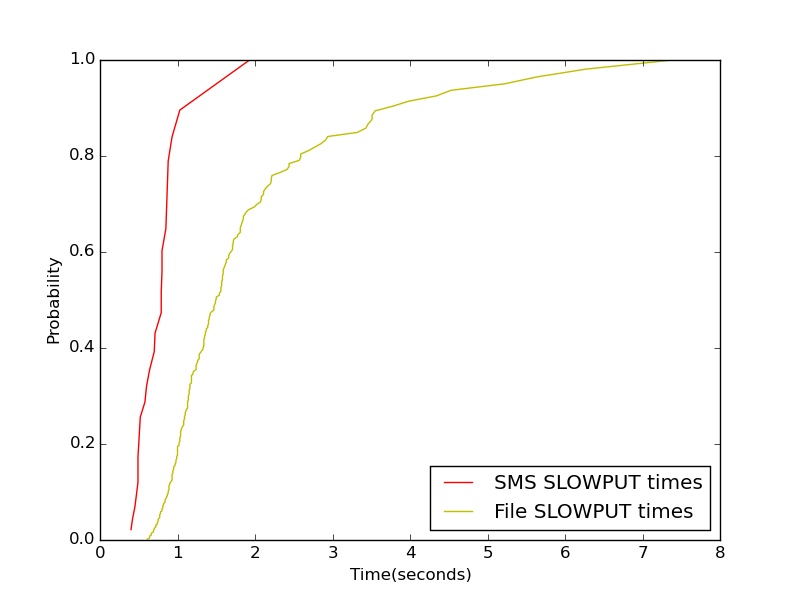}
	\caption{}
	\label{_cdf}
\end{subfigure}
\begin{subfigure}[b]{0.23\textwidth}
	\includegraphics[scale=0.16]{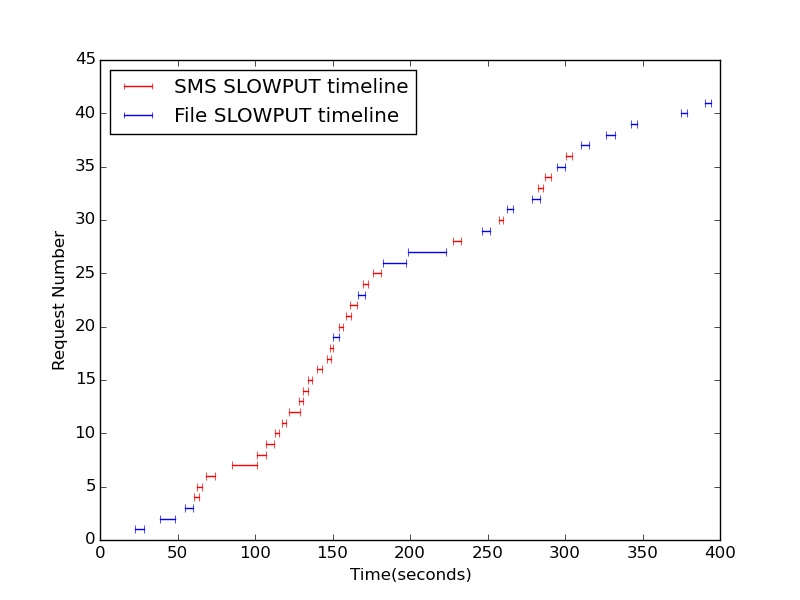}
	\caption{}
	\label{_queue}	
\end{subfigure}
\caption{CDF of time taken for the \texttt{SLOWPUT} of file vs SMS and a view of the queue in the P2P Marketplace application}
\label{smsvsfile}
\end{figure}
%
%
%

We report the performance of the P2P transactions application in
Figure~\ref{cdf}. The graph above shows the end-to-end latency between
the local server and cloud controller for the
\texttt{BUY},\texttt{SELL} and \texttt{SEARCH} operations. It is
important to recall that the \texttt{SELL} operation translates to the
\texttt{SLOWPUT} primitive, while \texttt{BUY} and \texttt{SEARCH}
translate to \texttt{FASTGET} and \texttt{FASTSEARCH} primitives
respectively. It can be observed in Figure~\ref{ghana_cdf} that
\texttt{BUY},\texttt{SELL} and \texttt{SEARCH} operations cause nearly
the same end-to-end latency. However in Figure~\ref{ny_cdf}, there is
a clear distinction between the time it takes for \texttt{SELL},
compared to \texttt{BUY} and \texttt{SEARCH}. This is expected because
\texttt{SELL} requests are queued and processed in a lazy manner. This
distinction is not visible in Figure~\ref{ghana_cdf} as that the
bandwidth between cloud and local server is low while the New York VCN
was connected to the cloud over a high bandwidth Ethernet link. This
results in similar latency measurements for all three operations. It
is important to note that no failures were experienced and all
messages sent from both nodes were received at the server despite the
fact that Ghana node was using a signal booster based backhaul. At a
user level user got SMS back for all the functions they
performed. They got SMS for \texttt{SELL} immediately because the
local component of the application gives the response immediately and
then executes the \texttt{SLOWPUT} operation while the response for \texttt{BUY} operation is sent only after response from cloud to maintain consistency. It is important to
note that even though there were no failures the users in Ghana
experienced high latency as is evident from the time given on x-axis
in Figure~\ref{ghana_cdf} and Figure~\ref{ny_cdf}

{\bf Sensing and Voice Applications:} File upload is a common
operation of the voice-based or sensing-based
applications. Figure~\ref{_cdf} shows the CDF of time it takes for
different \texttt{SLOWPUT} requests on the New York-based
VCN. Figure~\ref{_queue} shows the queue and how requests are
processed for the \texttt{SLOWPUT} primitive for the VCN in
Ghana. Figure~\ref{smsvsfile} shows that SMS-based applications will
work faster than the file based applications as they are smaller than
files. It also shows that file based applications can slow down the
SMS-based applications. This is true as both are placed in a common
queue for processing, and servicing larger files is more time
consuming than servicing smaller SMSs.

{\bf Summary of User Experiences:} The P2P marketplace application has
been tested with cohorts of farmers and traders from the local
community to perform BUY/SELL transactions. The distributed sensing
application has been used extensively to map farming boundaries and
collect meta-data about farms in the rural region. Using this
application, we have successfully built GIS agricultural maps of
different agricultural produce in the locality. The IVR based social
media application was inspired by the success of voice based citizen
journalism platforms like Polly~\cite{polly} and shows how one can
build such voice based applications to function effectively in rural
contexts using intermittency-aware VCNs.

%% file: content-experiences/conclusion.tex
\section{Conclusions}
This paper presents our experience in the design, implementation and
deployment of GreenLinks, a new ground-up platform that enables
intermittency-aware, reliable and available cellular services and
application support in rural contexts under extreme operational
environments with limited power. Greenlinks provides new primitives
for identity management and building new distributed applications to
operate effectively in highly intermittent environments. We present
our experiences building three example intermittency-aware distributed
applications and have used these applications on the ground in a rural
context in Ghana. In addition, to address operational challenges in
Ghana, our implementation supports mechanisms for building incrementally
deployable reliable backhauls and easy detection of whitespaces using
participatory sensing.